\documentclass{article}
\usepackage{graphicx}
\def\gtap{\ \raisebox{-.4ex}{\rlap{$\sim$}} \raisebox{.4ex}{$>$}\ }
\def\ms{\ifmmode{\overline{\rm MS}} \else{$\overline{\rm MS}$} \fi}

\begin{document}


\begin{titlepage}

\begin{flushright}
 UAB-FT-544       \\ 
 YNU-HEPTh-03-101 \\
 BUTP-2003/8      \\
 TU-687           \\
  hep-ph/0305063
\end{flushright}
\vspace{0.5cm}
\begin{center}
{\Large
{\bf Towards an exact evaluation of the supersymmetric}          \\[1.1ex]
{\bf $O(\alpha_s \tan \!\beta)$ corrections to
     $\overline{B} \to X_s \gamma$}}                             \\[5ex]
{\large  Francesca~Borzumati$^{\,a,b}$, Christoph Greub$^c$,
         and Youichi Yamada$^d$}                                 \\[3ex]
$^a${\it IFAE, Universitat Aut\`onoma de Barcelona, 
         08193 Bellaterra (Barcelona), Spain}                    \\
$^b${\it Dept. of Physics, Yokohama National University,
         Yokohama 240-8501, Japan}                               \\
$^c${\it Institut f\"ur Theoretische Physik, Universit\"at Bern,
         CH-3012 Bern, Switzerland}                              \\
$^d${\it Department of Physics, Tohoku University,
         Sendai 980-8578, Japan}
\end{center}
\vspace{1cm}
\begin{center} 
ABSTRACT \\
\vspace*{5mm}
\parbox{15cm}{The charged-Higgs contributions to the decay 
$\overline{B}\to X_s\gamma$ are discussed in the minimal supersymmetric
standard model at large $\tan\!\beta$. These contributions receive
two-loop ${\cal O}(\alpha_s\tan\!\beta)$ corrections by squark-gluino
subloops, which are nondecoupling in the limit of heavy superpartners
and possibly large. Their leading parts are already known and were
evaluated by using an effective two-Higgs-doublet Lagrangian.
Subleading corrections coming from higher-dimension operators in the
effective Lagrangian were ignored, although this is not, a priori,
justified when $m_{H^\pm}$ is not much smaller than the typical
supersymmetric mass $M_{\rm SUSY}$.  Here, we calculate all subleading
terms of the ${\cal O}(\alpha_s\tan\!\beta)$ corrections up to 
${\cal O}((m_t^2,m_{H^\pm}^2/M_{\rm SUSY}^2)^2)$, as well as all the exact 
two-loop diagrams with squark-gluino subloops, beyond the
effective-Lagrangian approximation.  Comments are made on the size of
these corrections.
}
\end{center}
\vfill 
{\begin{center} 
{\it Based on talks given by
 Y.~Yamada at ``KEK Theory Meeting on Collider Physics'', KEK, 
 Tsukuba, Japan, February 20-22, 2003, and 
 ``The 8th Accelerator and Particle Physics Institute (APPI2003)'',
  Appi, Iwate, Japan, February 25-28, 2003}
\end{center}}
\vfill
\end{titlepage}
\thispagestyle{empty}


\begin{center}
{\Large \bf Towards an exact evaluation of the supersymmetric    \\[1.01ex]
            $O(\alpha_s \tan \!\beta)$ corrections to
            $\overline{B}\to X_s \gamma$}                        \\ 
\vspace{10mm}
{\large Francesca Borzumati$^{a,b}$, Christoph Greub$^c$,
         and Youichi Yamada$^d$}                                 \\
\vspace{6mm}
\begin{tabular}{c}
$^a${\it IFAE, Universitat Aut\`onoma de Barcelona, 
         08193 Bellaterra (Barcelona), Spain}                    \\
$^b${\it Dept. of Physics, Yokohama National University,
         Yokohama 240-8501, Japan}                               \\
$^c${\it Institut f\"ur Theoretische Physik, Universit\"at Bern,
         CH-3012 Bern, Switzerland}                              \\
$^d${\it Department of Physics, Tohoku University,
          Sendai 980-8578, Japan}
\end{tabular}
\vspace{5mm}
\begin{abstract}
\baselineskip=15pt
The charged-Higgs contributions to the decay 
$\overline{B}\to X_s\gamma$ are discussed in the minimal supersymmetric
standard model at large $\tan\!\beta$. These contributions receive
two-loop ${\cal O}(\alpha_s\tan\!\beta)$ corrections by squark-gluino
subloops, which are nondecoupling in the limit of heavy superpartners
and possibly large. Their leading parts are already known and were
evaluated by using an effective two-Higgs-doublet Lagrangian.
Subleading corrections coming from higher-dimension operators in the
effective Lagrangian were ignored, although this is not, a priori,
justified when $m_{H^\pm}$ is not much smaller than the typical
supersymmetric mass $M_{\rm SUSY}$.  Here, we calculate all subleading
terms of the ${\cal O}(\alpha_s\tan\!\beta)$ corrections up to 
${\cal O}((m_t^2,m_{H^\pm}^2/M_{\rm SUSY}^2)^2)$, as well as all the 
exact two-loop diagrams with squark-gluino subloops, beyond the
effective-Lagrangian approximation.  Comments are made on the size of
these corrections.
\end{abstract}
\end{center}

\setlength{\parskip}{1.01ex}

\section{Introduction}
The inclusive width of the radiative decays of the $B$ mesons,
$\overline{B}\to X_s \gamma$, is well described by the short-distant
processes $b\to s\gamma$ and $b \to sg$, since nonperturbative
hadronic corrections are small and well under control. The partonic
processes have been evaluated up to the next-to-leading order in QCD.
See Ref.~\cite{BM} for a review listing the steps that brought to this
achievement, after it was noted that QCD plays a particularly
important role for the decay 
$\overline{B}\to X_s \gamma$~\cite{bsgQCD}.  The Standard Model (SM)
prediction for the branching ratio 
${\rm BR}(\overline{B}\rightarrow X_s \gamma)$ is, up to 
today~\cite{BGS},
\begin{equation}
 {\rm BR}(\overline{B}\rightarrow X_s \gamma)({\rm SM})=
 (3.54\pm 0.49) \times 10^{-4}\,.
\label{brsm}
\end{equation}
This average includes results obtained in the various papers of
Ref.~\cite{SMresult}.  The comparison of the SM result with the world
average~\cite{world} of the inclusive branching ratio from recent
experiments at Belle~\cite{Belle}, CLEO~\cite{CLEO}, and
BABAR~\cite{BABAR} detectors,
\begin{equation}
 {\rm BR}(\overline{B}\rightarrow X_s \gamma)({\rm exp.})=
(3.34\pm 0.38 )\times 10^{-4} \,,  
\label{brexp}
\end{equation}
is rather satisfactory.  In the SM, the process $b\to s\gamma$, as well 
as $b \to s g$, occurs through loops with $W^{\pm}$ and $t$-quark 
exchange, i.e. at the same level in perturbation theory at which
physics beyond the SM may contribute. The agreement between
experimental and SM results for 
${\rm BR}(\overline{B}\rightarrow X_s \gamma)$, therefore, is already 
good enough to constrain exotic
contributions.  Indeed, it is already routinely used to check whether
particular directions of parameter space of supersymmetric extensions
of the SM are viable or not.

In the minimal supersymmetric standard model (MSSM), new loop
contributions to the two radiative decays of the $B$ meson, $b\to s g$
and $b \to s \gamma$, come~\cite{BERTOLINI} from the charged-Higgs
boson, charginos, gluino and neutralino. Their contributions are often
comparable to or even larger than the SM one.  The inclusion of QCD
corrections to these contributions~\cite{QCDtoSUSYbsg}, however, is far
from having reached the level of precision already achieved in the SM,
at least for generic regions of the supersymmetric parameter space.  The
latest development in this direction has been the observation that the
SUSY-QCD corrections by squark-gluino subloops can be significant when
$\tan\!\beta$ is very large~\cite{effVbsg1,effVbsg2}.

In this talk, we focus on the contribution of the charged-Higgs boson
$H^\pm$ and analyze these two-loop SUSY-QCD corrections.  So far, only the
nondecoupling parts of these corrections have been evaluated, by using
an effective two-Higgs-doublet (2HD) Lagrangian where squarks and gluino
are integrated out~\cite{effVbsg1,effVbsg2}.  Subleading parts of these 
corrections, i.e. of order $((m_t^2,m_{H^\pm}^2)/M_{\rm SUSY}^2)^n$, 
generated by higher-dimensional operators in the effective Lagrangian,
were omitted in these studies, although they could give potentially
large contributions when $m_{H^\pm}$ is nonnegligible with respect to the
typical supersymmetric mass $M_{\rm SUSY}$.

Here, we calculate some of these subleading corrections, i.e.  up to
${\cal O}((m_t^2,m_{H^\pm}^2/M_{\rm SUSY}^2)^2)$. We also evaluate exactly
all two-loop diagrams correcting at ${\cal O}(\alpha_s \tan \!\beta)$
the charged-Higgs contribution to $b \to s \gamma$ and $b\to s g$. This
calculation, clearly, encompasses the effective-Lagrangian approximation
and includes all subleading terms 
$((m_t^2,m_{H^\pm}^2)/M_{\rm SUSY}^2)^n$.  In this talk, after discussing
the effective-Lagrangian formalism in Section~\ref{EffLagForm}, the
charged-Higgs contribution to $b \to s \gamma$ and $b\to s g$ in 
Section~\ref{RadDecContrib}, we show in Section~\ref{NumRes} numerical
results for the charged-Higgs contributions to the Wilson coefficients
$C_7$ and $C_8$ at the matching scale, $\sim M_W$.  The evaluation of
these same coefficients at low scale, $\sim m_b$, and therefore of the
${\rm BR}(\overline{B}\to X_s \gamma)$, becomes, at this point,
straightforward.  We have, however, refrained from showing results for
${\rm BR}(\overline{B}\to X_s \gamma)$ and from extracting exclusion
plots for the charged-Higgs boson mass, since the exact calculation of 
the $O(\alpha_s \tan \!\beta)$ corrections to the the $W^\pm$ 
contribution, or at least the calculation of the subleading 
${\cal O}(m_t^2/M_{\rm SUSY}^2)$ and 
${\cal O}((m_t^2/M_{\rm SUSY}^2)^2)$ terms of these corrections, is 
not yet available.  We comment on the outcome of our calculation in
Section~\ref{NumRes} and we conclude in Section~\ref{concl}.

\section{$H^\pm$ couplings to quarks}
\label{EffLagForm}
\subsection{Tree-level couplings} 
\label{EffLagtree}
The MSSM has two Higgs doublets $H_D$ and $H_U$, $H_D = (H_D^0, H_D^-)
$ and $H_U = (H_U^+, H_U^0) $, which break the SU(2)$\times$U(1) gauge
symmetry through the vacuum expectation values (VEVs) of their neutral
components.  The two VEVs are related to the $W$-boson mass as 
$M_W^2 = g_2^2\bar{v}^2/2\equiv g_2^2(v_D^2+v_U^2)/2$. Their ratio is
conventionally called $\tan\! \beta$, $\tan\!\beta\equiv v_U/v_D$.
They form the following mass eigenstates: two CP-even scalars, 
($h^0$, $H^0$); one CP-odd pseudoscalar, $A^0$; the two states of a 
charged-Higgs boson, $H^{\pm}$; and the unphysical Nambu-Goldstone 
modes, ($G^{\pm}$, $G^0$). The charged scalars ($H^{\pm}$, $G^{\pm}$) 
are related to the charged components of the gauge eigenfields 
$H^{\pm}_{D,U}$ as
\begin{equation}
\left( \begin{array}{c} G^\pm \\ H^\pm \end{array} \right) =
\left( \begin{array}{rr} \cos\!\beta & -\sin\!\beta \\ 
                         \sin\!\beta &  \cos\!\beta 
       \end{array} \right) 
\left( \begin{array}{c} H^\pm_D \\ H^\pm_U \end{array} \right)\, .
\end{equation}

At the tree-level, the couplings of the Higgs doublets $H_i(i=D,U)$ 
to quarks  
obey the selection rule of the 2HD model of Type~II. That is to say, 
$(d_{R})_i=(d,s,b)_R$ couple only to $H_D$, whereas 
$(u_{R})_i=(u,c,t)_R$
couple only to $H_U$, as shown by the interaction Lagrangian in the
gauge eigenbasis:
\begin{equation}
{\cal L} \ = \  
 \bar{d}_{R} \, Y^d q_{L} \cdot H_D 
\ - \,
 \bar{u}_{R} \, Y^u q_{L} \cdot H_U  
\ + \ ({\rm h.c.}), 
\label{lagtree}
\end{equation}
in which the SU(2)-invariant multiplication of doublets was adopted
($A\cdot B \equiv \epsilon_{ij} A_i B_j$, with 
$\epsilon_{12} = -\epsilon_{21}=1$ and 
$\epsilon_{11}= \epsilon_{22}=0$).  This constraint is a consequence 
of supersymmetry.

After diagonalization of the Yukawa matrices $Y^q$, 
$Y^q \rightarrow {\rm diag}(h_{q})$, and the breaking of 
SU(2)$\times$U(1), the $b$- and $t$-quark acquire masses 
\begin{equation}
  m_b \,= \,h_b v_D \,= \,h_b\bar{v}\cos\!\beta\,, 
\hspace*{0.8truecm}
  m_t \,= \,h_t v_U \,= \,h_t\bar{v}\sin\!\beta\,, 
\label{mqtree}
\end{equation}
and the $H^+$ couplings to the $t$- and $b$-quarks become
\begin{equation}
{\cal L}\ = \ 
 V_{tb} h_b\sin\!\beta \, H^+ \bar{t}_L b_R  +
 V_{tb} h_t\cos\!\beta \, H^+ \bar{t}_R b_L   
\ +  ({\rm h.c.})\,.
\label{Hplagtree}
\end{equation}
Here, $V_{tb}$ is a element of the CKM matrix. Of the two 
couplings
\begin{eqnarray}
g(H^+ \bar{t}_L b_R) 
& = & 
 V_{tb} h_b\sin\!\beta \ = \ V_{tb} \frac{m_b}{\bar{v}}\tan\!\beta
\nonumber \\ 
g(H^+ \bar{t}_R b_L) 
& = & 
 V_{tb} h_t\cos\!\beta \ = \ V_{tb} \frac{m_t}{\bar{v}}\cot\!\beta
\label{couplingstree}
\end{eqnarray}
the first is greatly enhanced for large $\tan\!\beta$.

\subsection{Couplings up to ${\cal O}(\alpha_s \tan\! \beta)$ }
\label{EffLagloop}
\begin{figure}[ht] 
\vspace{0.3truecm}
\begin{center} 
\includegraphics[width= 4cm]{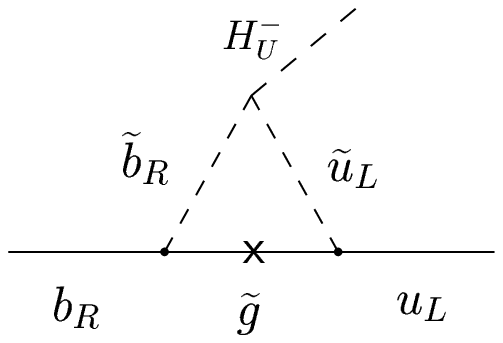} \hspace{5mm}
\includegraphics[width= 4cm]{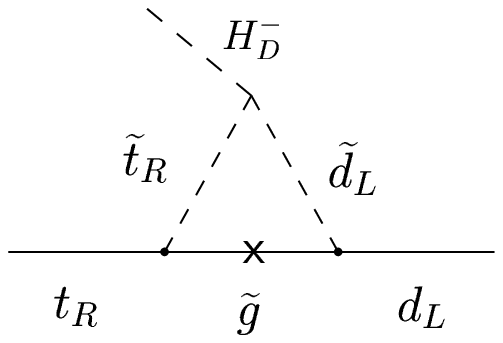}
\end{center} 
\vspace{-0.3truecm}
\caption[f1]{$\bar{b}_R u_L H_U^-$ and $\bar{d}_L t_R H_D^-$ vertices
 generated by squark-gluino loop, shown in the gauge eigenbasis of
 squarks.}
\label{EVertices} 
\end{figure}

After supersymmetry breaking, loop corrections depending on soft
supersymmetry-breaking parameters generate effective couplings of
quarks $d_{iR}$ to $H_U^\dagger$ and of $u_{iR}$ to $H_D^\dagger$,
which are forbidden at the tree-level. Squark and gluino loops
inducing the couplings $\bar{b}_R u_L H_U^-$ and 
$\bar{d}_L t_R H_D^-$, are shown explicitly in Fig.~\ref{EVertices}.

At momentum scales sufficiently smaller than the typical
supersymmetric mass $M_{\rm SUSY}$, these contributions can be
expressed in terms of an effective 2HD Lagrangian (not of Type~II
anymore)
\begin{equation}
{\cal L}^{\rm eff}\ = \ 
 \bar{d}_R Y^d \, q_L \cdot \!(H_D -\!\Delta_{d_R,q} H_U^c )
\ - \ 
 \bar{u}_R Y^u \, q_L \cdot \!(H_U +\!\Delta_{u_R,q} H_D^c)
\ + \ ({\rm h.c.})\,, 
\label{lagloop}
\end{equation}
where $H_D^c=(-H_D^+, H_D^{0*})$ and $H_U^c=(-H_U^{0*}, H_U^-)$,
respectively, and where the index $q$ in $\Delta_{d_R,q}$ and 
$\Delta_{u_R,q}$ is understood to denote a left-handed quark.  The 
effective couplings $\Delta_{d_R,q}$ and $\Delta_{u_R,q}$ induced by 
SUSY-QCD loops (as well as Higgsino-squarks loops) are of order
\begin{equation}
\Delta_{d_R,q}, \Delta_{u_R,q}   
= {\cal O}\left(\alpha_s 
                \frac{\,\mu \,m_{\tilde{g}}\,}{M_{\tilde{q}}^2}
          \right) 
= {\cal O}(\alpha_s M_{\rm SUSY}^0)\,. 
\end{equation}
i.e. do not decouple~\cite{dmb,carenaH0,carenaHp} in the limit of heavy
superpartners, $M_{\rm SUSY}\to\infty$. Although sufficiently smaller
than unity, they can induce large corrections to the $H^{\pm}$
couplings to quarks for large $\tan\!\beta$.

After the breaking of SU(2)$\times$U(1), masses and $H^\pm$-couplings 
of the $t$- and $b$-quarks become, at this order,
\begin{eqnarray}
{\cal L}_{\rm (mass,int)}^{\rm eff} 
& \!\!\!= \!\!\! & 
  -
h_b\bar{v}\cos\!\beta
\left[1+\!\Delta_{b_R,b} \tan\!\beta\right] \bar{b}b 
\ - 
h_t\bar{v}\sin\!\beta
\left[1+\!\Delta_{t_R,t} \cot\!\beta\right] \bar{t}t 
\nonumber \\
&   &
  +
V_{tb} h_b\sin\!\beta 
\left[1-\!\Delta_{b_R,t} \cot\!\beta\right] H^+ \bar{t}_L b_R
\ + 
V_{tb} h_t\cos\!\beta
\left[1-\!\Delta_{t_R,b} \tan\!\beta\right] H^+ \bar{t}_R b_L 
\ +\! \left({\rm h.c.}\right) \,,
\label{lagloop2}
\end{eqnarray} 
where CP-violating phases in supersymmetric parameters were dropped
for simplicity. Notice that the $\Delta_{b_R,q}$ and $\Delta_{t_R,q}$
of Eq.~(\ref{lagloop}) are now split into $\Delta_{b_R,b}$, 
$\Delta_{b_R,t}$ and $\Delta_{t_R,b}$, $\Delta_{t_R,t}$. The elements
of each of these two pairs differ by SU(2)$\times$U(1) breaking effects. 
This splitting is achieved by including in the effective lagrangian 
of Eq.~(\ref{lagloop}) additional higher-dimension operators. See 
for example $\bar{b}_R (q_L \cdot H_U^c) H_U^{\dagger} H_U$ and 
$\bar{b}_R (q_L \cdot H_D) H_U^{\dagger} H_U$.

The first line of Eq.~(\ref{lagloop2}) denotes the running $b$-quark 
mass within the SM, that is to say:
\begin{equation}
 m_b({\rm SM}) \ = \ h_b\bar{v}\cos\!\beta
                 \left[1+\Delta_{b_R,b} \tan\!\beta \right] \,.
\label{dmbeq}
\end{equation} 
The corrections $\Delta_{b_R,b} \tan \!\beta$, of 
${\cal O}(\alpha_s\tan\!\beta)$ are potentially large~\cite{dmb} 
for $\tan \! \beta \gtap (\Delta_{b_R,b})^{-1}$.  As a result, when
parametrized by $m_b({\rm SM})$, the coupling $H^+\bar{t}_L b_R$ may
significantly deviate~\cite{carenaH0,carenaHp} from the 
(renormalization group improved) tree-level result, 
\begin{equation}
g\left(H^+ \bar{t}_L b_R\right)({\rm eff}) 
\ \sim \ 
V_{tb} h_b \sin\!\beta
\ = \ 
V_{tb}\frac{m_b({\rm SM})}{\bar{v}}
\frac{\tan\!\beta}{1+\Delta_{b_R,b} \!\tan\beta } \,.
\label{Hpdmb}
\end{equation}
Note that Eq.~(\ref{Hpdmb}) incorporates the resummation of all
$(\alpha_s\tan\!\beta)^n$ corrections to $m_b$~\cite{carenaHp}.

The corrections to the vertex $H^+\bar{t}_R d_L$ are factors 
$\Delta_{t_R,d} \tan\!\beta$ and also potentially large. The relevant loop 
diagram is shown on the right side of Fig.~\ref{EVertices}. These
corrections give rise to effective couplings 
\begin{equation}
g\left(H^+ \bar{t}_R d_L\right)({\rm eff}) 
\ \sim \ 
V_{td} \frac{m_t({\rm SM})}{\bar{v}}
        \left( \cot \! \beta - \!\Delta_{t_R,d}\right)\,,
\label{Hpvert}
\end{equation}
where the second term in parentheses can very well be of the same 
order of magnitude of the first.

For momentum scales nonnegligible with respect to the superpartner
masses, the above effective Lagrangian has to be enlarged to incorporate
additional higher-dimensional operators that have momentum dependence.
\begin{figure}[ht] 
\vspace{0.3truecm}
\begin{center} 
\includegraphics[width= 4cm]{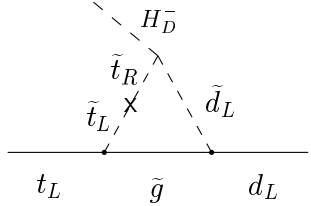}
\end{center} 
\vspace{-0.3truecm}
\caption[f1]{$\bar{d}_L t_L H_D^-$ vertices generated by squark-gluino
 loops, shown in the gauge basis of squarks.}
\label{newEVertex} 
\end{figure} 
For example, the diagram in Fig.~\ref{newEVertex} generates effective 
operators of dimension {\it six}, such as 
$ \bar{q}_L H_D \gamma^\mu (\partial_\mu q_L)\cdot H_U$, 
and in particular, 
$H_D^- \bar{d}_L \gamma^\mu (\partial_\mu t_L) H_U^0$, where $H_U^0$ 
gets a vacuum expectation value.  This gives rise to a new set of 
couplings $H^+\bar{t}_L \gamma^\mu d_L$, which in the large 
$\tan \! \beta$ limit are of size 
\begin{equation}
g\left(H^+(q) \,\bar{t}_L(p) \gamma^\mu d_L(q\!-\!p)\right)({\rm eff}) 
\ \sim  \ 
V_{td} \,p^\mu \frac{m_t({\rm SM})}{M_{\rm SUSY}^2} 
        \,\Delta_{t_L,d}\sin\!\beta 
\ \sim  \ 
V_{td} \,p^\mu \frac{m_t({\rm SM})}{M_{\rm SUSY}^2} 
        \,\Delta_{t_L,d} \,.
\label{Hpvertnew}
\end{equation}
These operators are suppressed by inverse powers of $M_{\rm SUSY}$ and
decouple in the limit of heavy superpartners.

\section{$H^+$ contributions to $b \to s \gamma$ and $b \to s g$ up to
 ${\cal O}(\alpha_s\tan\!\beta)$ } 
\label{RadDecContrib}
\begin{figure}[ht] 
\vspace{0.3truecm}
\begin{center} 
\includegraphics[width= 6.3cm]{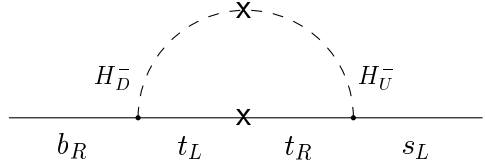}
\end{center} 
\vspace{-0.3truecm}
\caption[f1]{\small $b\to s\gamma$ by charged-Higgs exchange, with
 helicity flip on the internal fermion line. The Higgs doublet
 constituent of the charged-Higgs state is indicated explicitly.}
\label{CH-LO} 
\end{figure}

It is well known that $b\to s \gamma$ and $b \to s g$ are very sensitive
to $\tan\!\beta$~\cite{bsgtanbeta}, at times even dangerously
so~\cite{bsgdmb}. However, at the one-loop level, the charged-Higgs
contributions to these decays remains fairly independent from this
parameter, for $\tan \!\beta\gtap 3$. In this regime, the dominant part
of the $H^{\pm}$ contribution to the decays $b\to s\gamma$ and 
$b \to s g$, comes from the diagrams in Fig.~\ref{CH-LO}, where the
photon and gluon have still to be attached in all possible ways. In this
Figure, it is shown explicitly how the factor $\tan\!\beta$ for the
vertex $\bar{t}_Lb_RH_D^+$ is cancelled by the mixing between $H_D$ and
$H_U$, which brings in a suppression factor 
$\sin 2 \beta \sim 1/\tan\!\beta$.  Phrased in an equivalent way, the
decay amplitude for this diagram becomes insensitive to $\tan\!\beta$,
due to the cancellation of the factor $\tan\!\beta$ between the
$\bar{t}_Lb_RH^+$ and $\bar{s}_Lt_RH^-$ vertices.

\begin{figure}[ht] 
\vspace{0.3truecm}
\begin{center} 
\includegraphics[width= 6.3cm]{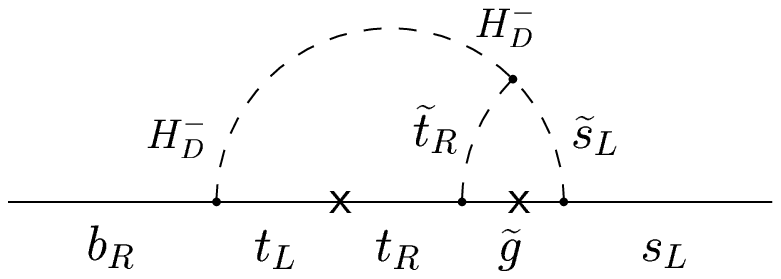} \hspace{5mm}
\includegraphics[width= 6.3cm]{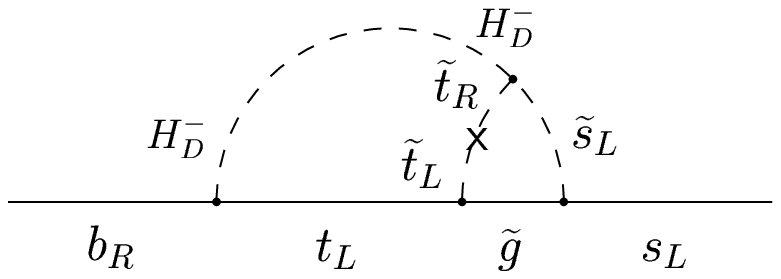}
\end{center} 
\vspace{-0.3truecm}
\caption[f1]{Gluino corrections to the $H^-\bar{s}_Lt_R$ and
 $H^-\bar{{s}_L}t_L$ vertices contributing to the decays
 $b\to s\gamma$ and  $b\to sg$. The photon can be attached to the
 $t$, $\tilde{t}$, $\tilde{s}_L$, and $H^-$ lines, the
 gluon to the $t$, $\tilde{t}$, $\tilde{s}_L$, and
 $\tilde{g}$ lines.}
\label{CH-NLO} 
\end{figure}

However, since the vertex $H_D^-\bar{s}_Lt_R$ is generated at the 
one-loop level, it is possible to avoid this $\tan \!\beta$
cancellation, when considering two-loop contributions to $b\to s\gamma$ 
and $b \to s g$.  The corresponding diagram, where again, the photon
and the gluon are still to be attached in all possible ways, is shown
on the left of Fig.~\ref{CH-NLO}.  Since also the vertex 
$H_D^-\bar{s}_Lt_L$ gets generated at the one-loop level, there is 
another diagram in which it is possible to avoid the cancellation of 
$\tan\!\beta$ of the lowest order contribution. This is shown on the
right side of Fig.~\ref{CH-NLO}. Note that, being the vertex 
$H_D^-\bar{s}_Lt_L$ of decoupling type, also the corresponding 
$b\to s \gamma$ and $b \to s g$ contributions decouple in the limit
$M_{\rm SUSY} \to \infty$.  Both classes of contributions coming from
the two diagrams in Fig.~\ref{CH-NLO} are, nevertheless, of 
${\cal O}(\alpha_s\tan\!\beta)$ with respect to the lowest order
contribution. They are potentially large, without invalidating
perturbation theory: their largeness derives from the suppression of
the lowest order term.

Overall, potentially large ${\cal O}(\alpha_s\tan\!\beta)$ corrections
to the charged-Higgs contribution to $b \to s \gamma$ (and $b \to s g$)
may come from:
\begin{enumerate}
\item 
 mass corrections, $\delta m_b$, to the vertex
  $H^+\bar{t}_Lb_R$~(\ref{Hpdmb}), see Ref.~\cite{bsgdmb}; 
\item 
 proper vertex corrections to the coupling
 $H^-\bar{s}_Lt_R$~\cite{effVbsg1,effVbsg2} as well as $H^-\bar{s}_Lt_L$,
 shown in Fig.~\ref{CH-NLO};
\item 
 corrections to the couplings 
 $H^-\bar{s}_Lt_R \gamma$ and  $H^-\bar{s}_Lt_L \gamma$, obtained from 
 the two diagrams in Fig.~\ref{CH-NLO} when the photon is attached 
 to the  $\tilde{t}$- or the $\tilde{s}$-squarks, and  
 corrections to the couplings  
 $H^-\bar{s}_Lt_R g$ and  $H^-\bar{s}_Lt_L g$, obtained from 
 the two diagrams in Fig.~\ref{CH-NLO} when the gluon is attached 
 to the  $\tilde{t}$- or the $\tilde{s}$-squarks, or the gluino.
\end{enumerate}

These corrections, induced by squarks and gluino subloops, have been so
far calculated by using the effective 2HD Lagrangian of
Eq.~(\ref{lagloop}), with all superpartners integrated out~\footnote{A
 similar procedure has been applied to
 collect higher order corrections of type ${\cal O}(\alpha \tan\!\beta)$
 in Ref.~\cite{DIAZ}.}. This approach is justified as far as all the
momenta of the fields in Eq.~(\ref{lagloop}) are sufficiently smaller
than the squarks and gluino. This condition is clearly satisfied for the
counterterm $\delta m_b$ and its contribution to the
$H^+\bar{t}_Lb_R$ vertex, see Eq.~(\ref{Hpdmb}).

For the corrections to the $H^-\bar{s}_Lt_R$ vertex, the validity of the
effective Lagrangian of Eq.~(\ref{lagloop}) requires $m_t$ and
$m_{H^\pm}$ to be smaller than $M_{\rm SUSY}$. The condition 
$m_{H^\pm}\ll M_{\rm SUSY}$, however, is often violated in the case of
well-known candidates for the mechanism of supersymmetry breaking. Thus,
in the resulting models, and in general when 
$m_{H^\pm}\gtap M_{\rm SUSY}$, it is possible that the amplitude of the
full two-loop diagrams deviates significantly from that obtained by
making use of the effective 2HD Lagrangian of Eq.~(\ref{lagloop}).
One may try to handle the case of $m_{H^\pm}$ nonnegligible with respect
to $M_{\rm SUSY}$ by including in the effective Lagrangian
higher-order operators, which are suppressed by inverse powers of
$M_{\rm SUSY}^2$.  This is equivalent to making
an expansion of the two-loop integrals corresponding to the diagrams 
in Fig.~\ref{CH-NLO} with respect to $m_{H^\pm}^2/M_{\rm SUSY}^2$ and
$m_t^2/M_{\rm SUSY}^2$. 
The operators inducing the vertex $H^-\bar{s}_Lt_L$
(see discussion at the end of Section~\ref{EffLagloop}), as well as
those inducing 
the vertices $H^-\bar{s}_Lt_R \gamma$, $H^-\bar{s}_Lt_L\gamma$, and
$H^-\bar{s}_Lt_R g$, $H^-\bar{s}_Lt_L g$, 
are already of decoupling type, i.e. suppressed by inverse powers of
$M_{\rm SUSY}^2$. Nevertheless, additional expansions of the 
two-loop integrals may still be needed.

A systematic way of making these expansions ad simultaneously 
keep into account all needed higher-dimensional operators is
provided by the Heavy Mass Expansion (HME)~\cite{HME}.  
Using this technique, we find for example that, among the operators 
of dimension {\it six}
to be added, it is necessary to include also the very same ${\cal O}_7$
and ${\cal O}_8$,
\begin{equation}
{\cal O}_7(\mu) = 
 \frac{e}{16\pi^2} m_b(\mu)\bar{s}_L\sigma^{\mu\nu} b_R F_{\mu\nu}\,, 
\hspace{0.8truecm}
{\cal O}_8(\mu) = 
 \frac{g_s}{16\pi^2}m_b(\mu)\bar{s}_L\sigma^{\mu\nu}T^a b_RG^a_{\mu\nu}\,, 
\label{O7and8}
\end{equation}
which will be part of the effective Hamiltonian 
used for the calculation of amplitudes of radiative $b$ decays. 
{}$F_{\mu\nu}$ and $G^a_{\mu\nu}$ in these operators are 
the field strengths of the photon and the gluon, respectively.  
Details on the use of the HME 
are given in Ref.~\cite{BGY}.

Through the HME technique we have evaluated terms up to  
${\cal O}((m_{H^\pm}^2,m_t^2/M_{\rm SUSY}^2)^2)$. Notice that the first
term of this expansion is, for small values of $m_{H^\pm}$, of the order
of magnitude of the SU(2)$\times$U(1)-breaking corrections to the
coefficients of the effective Lagrangian~(\ref{lagloop}); see
Ref.~\cite{Buras}.  For $m_{H^\pm}$ closer to or even larger than 
$M_{\rm SUSY}$, however, one may question even the validity of this
expansion. Of course, the truncation up to the 
${\cal O}((m_{H^\pm}^2,m_t^2/M_{\rm SUSY}^2)^2)$ can only be justified by
comparing with the result of the complete two-loop calculation, which
clearly goes beyond the effective-Lagrangian approach. We have,
therefore, also calculated all two-loop diagrams exactly, using methods
described in Ref.~\cite{GvdB}.
In both cases, i.e. whether we make use of an expansion or not, we need
to calculate the ${\cal O}(\alpha_s\tan\!\beta)$ terms arising from both
diagrams shown in Fig.~\ref{CH-NLO}.  Nondecoupling corrections in the
effective-Lagrangian approach come from the left diagram with
photon/gluon emitted only from the $t$-quark or charged-Higgs boson. All
other diagrams are decoupling in the $M_{\rm SUSY}\to\infty$ limit, 
and have not been included in previous 
studies~\cite{effVbsg1,effVbsg2,DAmbrosio,Buras}.

\begin{figure}[t] 
\vspace{0.3truecm}
\begin{center} 
\includegraphics[width= 7.9cm]{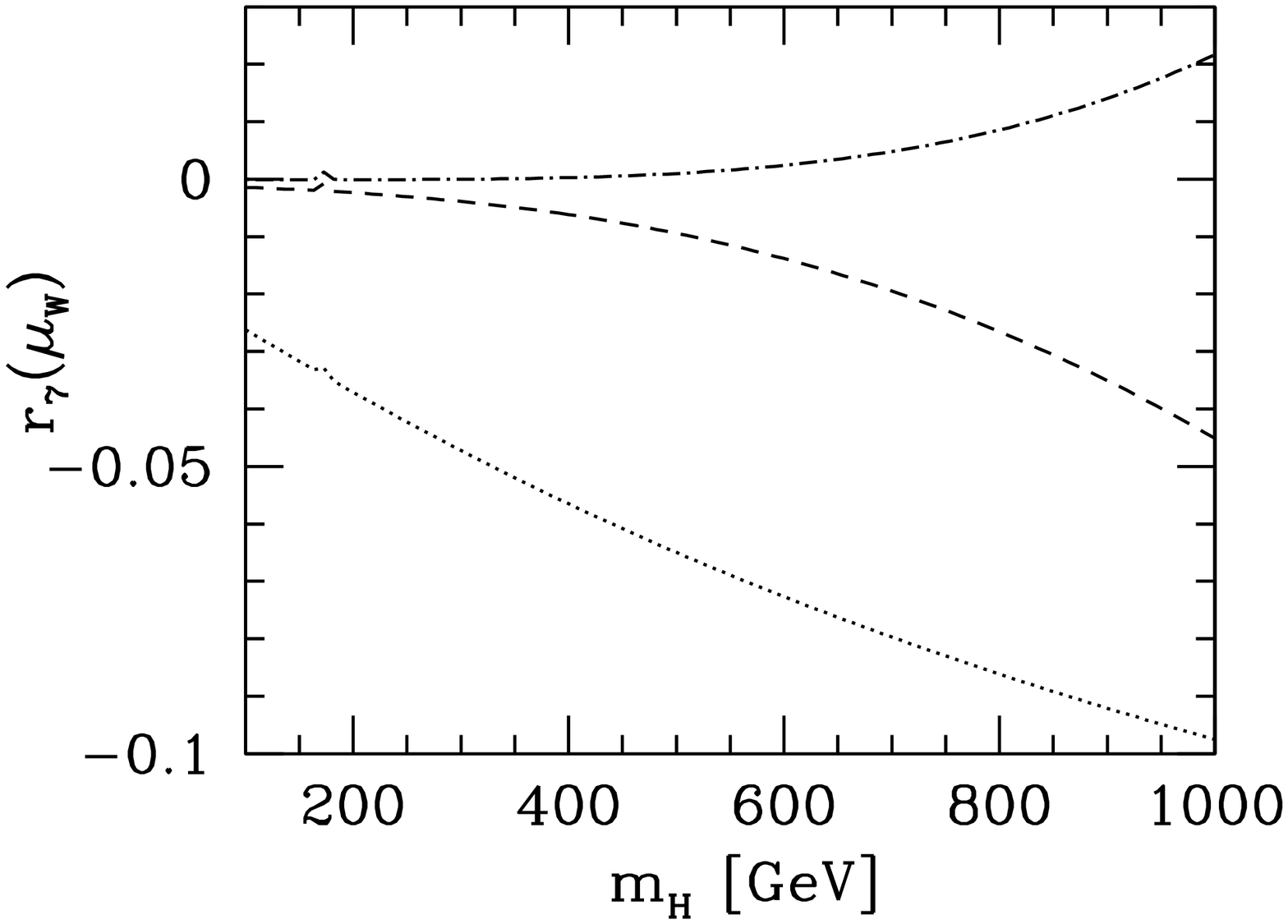}
\hspace{2mm}
\includegraphics[width= 7.9cm]{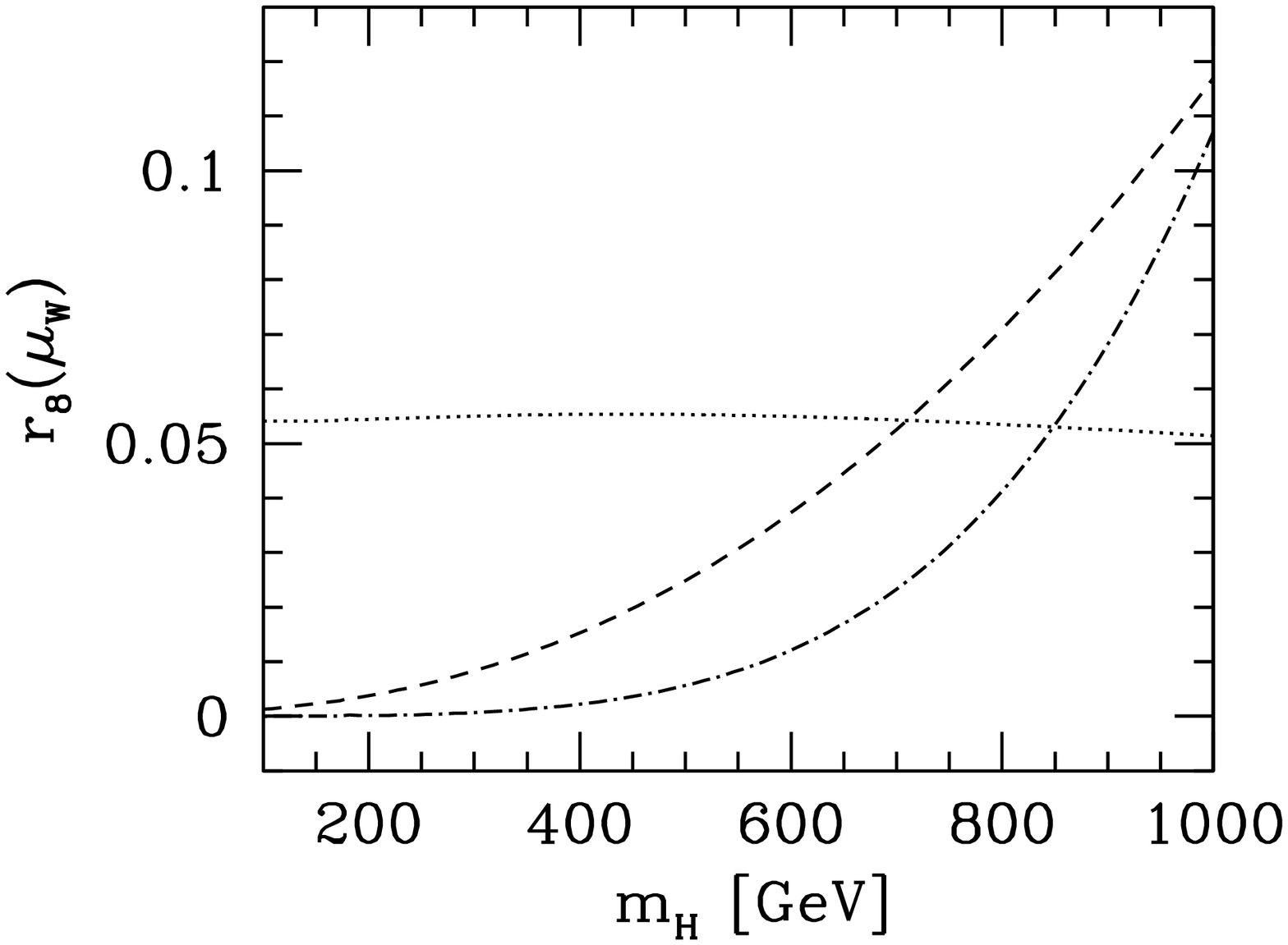} 
\end{center} 
\vspace{-0.5truecm}
\caption[f1]{\small Ratios $r_{7,8}(\mu_W)$, defined in
 Eq.~(\ref{plotratio}),  as functions of $m_{H^\pm}$. The supersymmetric
 spectrum considered here is the spectrum~I, given in the text. 
 The dotted lines show the goodness of the leading-order approximation
 of the two-loop calculation, the dashed and dot-dashed lines, the 
 goodness of the approximation in which the first and the second 
 subleading terms in $(m_{H^\pm}^2,m_{t}^2)/M_{\rm SUSY}^2$ are
 included.}
\label{c78figheavy} 
\vspace{5mm}
\begin{center} 
\includegraphics[width= 7.9cm]{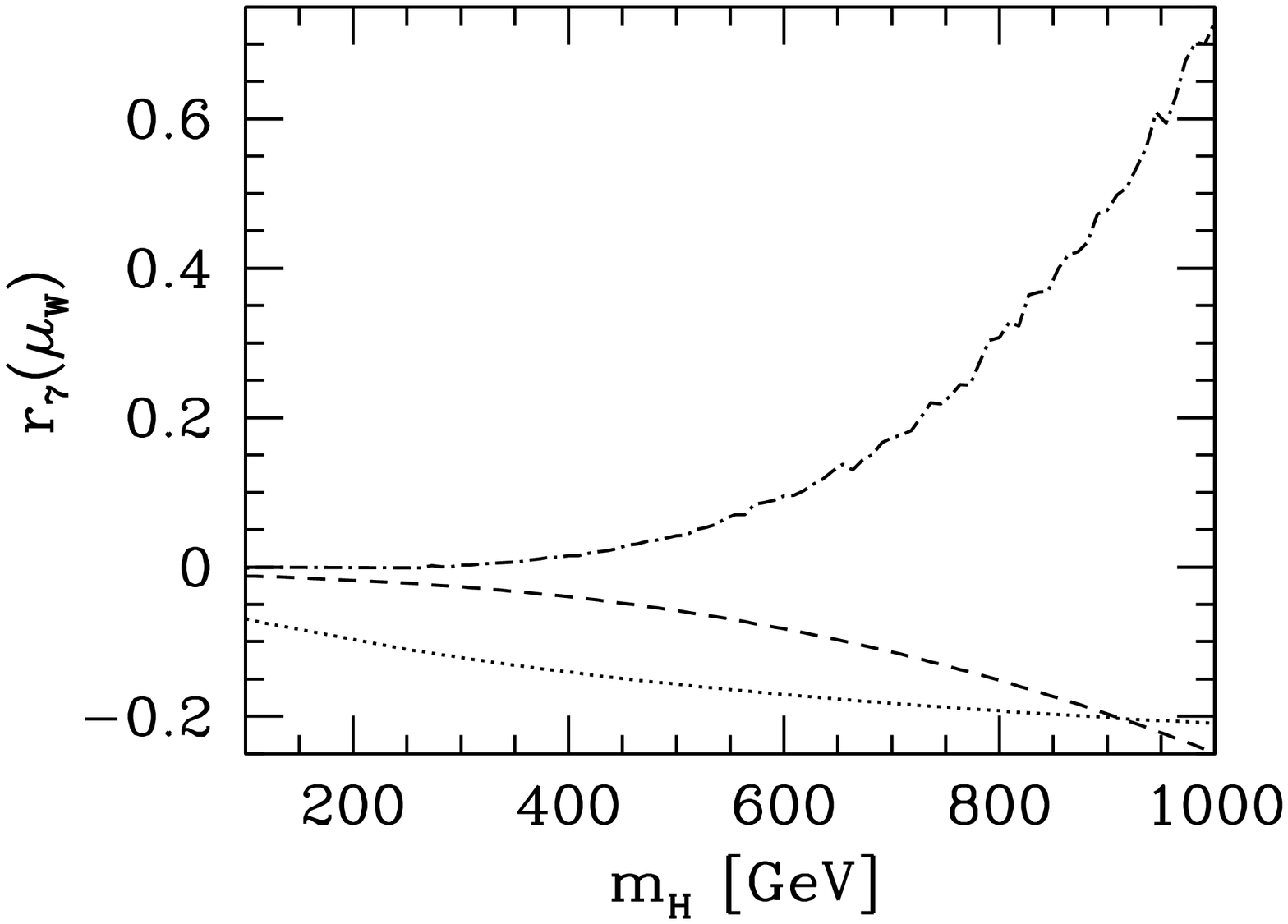}
\hspace{2mm}
\includegraphics[width= 7.9cm]{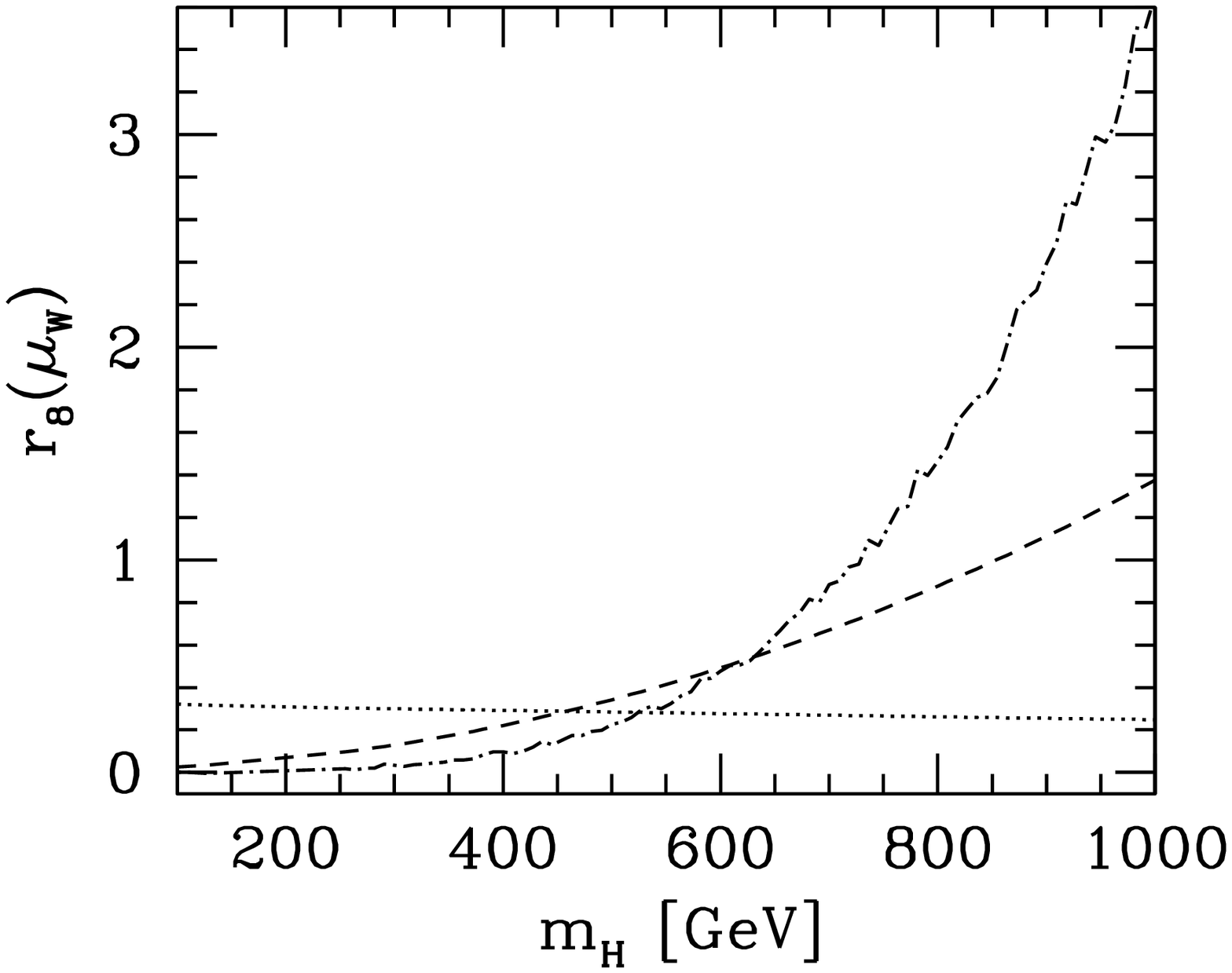} 
\end{center} 
\vspace{-0.5truecm}
\caption[f1]{\small Same as in Fig.~\ref{c78figheavy}, for the 
 spectrum~II.}
\label{c78figlight} 
\end{figure} 
\section{Numerical results}
\label{NumRes}
We present here results for the charged-Higgs contributions to the
Wilson coefficients $C_7(\mu_W)$ and $C_8(\mu_W)$ of the operators 
${\cal O}_7$ and ${\cal O}_8$ at the matching scale $\sim \mu_W$, 
which we have chosen to be $M_W$.  We incorporate the corrections to 
$C_7(\mu_W)$ and $C_8(\mu_W)$ discussed in the previous Section.  Our 
normalization of these Wilson coefficients is the conventional one, as
follows from the definition of the effective Hamiltonian used for the
calculation of amplitudes of radiative $b$ decays,
\begin{equation}
H_{\rm eff} \supset
 -\frac{4G_F}{\sqrt{2}}V^*_{ts}V_{tb}
 \left( C_7(\mu) {\cal O}_7(\mu) + C_8(\mu) {\cal O}_8(\mu) 
 \right) \,.
\end{equation}

We show results in which the two-loop diagrams in Fig.~\ref{CH-NLO} 
are calculated exactly, and results in which the off-shell vertices 
$H^- s_L t_R$ and $H^- s_L t_L$ are treated in an effective-Lagrangian
approach, with inclusion of leading and subleading terms up to overall
suppression factors $1/(M_{\rm SUSY}^2)^2$.  We compare these
approximated results, which we call 
$C_{7,8}(\mu_W)\vert_{\rm approx}$, with the exact result, denoted by
$C_{7,8}(\mu_W)\vert_{\rm exact}$.  To make this comparison more
transparent, we plot in Figs.~\ref{c78figheavy} and~\ref{c78figlight}
the ratios
\begin{equation}
r_i(\mu_W) \equiv 
\displaystyle{
  \frac{C_i(\mu_W)\vert_{\rm approx} -C_i(\mu_W)\vert_{\rm exact}}
       {C_i(\mu_W)\vert_{\rm exact}} } 
\hspace*{5truemm}
 (i = 7,8)\,,
\label{plotratio}
\end{equation}
as functions of $m_{H^\pm}$. In these ratios, the mass correction 
$\delta m_b$ to the vertex $H^+\bar{t_L} b_R$ cancels out. Therefore, 
we need to specify only a relatively small number of parameters.

For Fig.~\ref{c78figheavy}, we have chosen a heavier superpartner
spectrum, called here spectrum~I,
$(m_{\tilde{s}_L},m_{\tilde{t}_1},m_{\tilde{t}_2})=
 (700,500,450)$ GeV, the stop-sector mixing angle $\cos \theta_t = 0.8$, 
$\tan\!\beta=30$, $m_{\tilde{g}}=600$ GeV, and $\mu=550$ GeV, 
whereas for Fig.~\ref{c78figlight} a lighter spectrum is considered:
$(m_{\tilde{s}_L},m_{\tilde{t}_1},m_{\tilde{t}_2})=
 (350,400,320)$ GeV, $\cos \theta_t = 0.8$, 
$\tan\!\beta=30$, $m_{\tilde{g}}=300$ GeV, and $\mu=450$ GeV. This is 
denoted as spectrum~II. 
As for other input parameters, we have used 
$m_t(\mu_W) = 176.5\,$GeV, which corresponds to a pole mass of 
$175\,$GeV, and $\alpha_s(\mu_W) =0.12 $.

In both sets of Figures, the dotted lines denote the ratios $r_7(\mu_W)$
and $r_8(\mu_W)$ in which only the leading-order approximation is used
for the evaluation of the two-loop diagram, i.e. the nondecoupling
contribution of Ref.~\cite{effVbsg1}; the dashed (dot-dashed) lines
denote the ratios in which the first (second) subleading terms in
$(m_t^2,m_{H^\pm}^2)/M_{\rm SUSY}^2$ are also included.

These Figures show explicitly that for $m_{H^\pm}$ sufficiently 
smaller than $M_{\rm SUSY}$, the use of the HME, and therefore of 
the effective-Lagrangian approach, is indeed meaningful: 
the inclusion of an additional term of the expansion brings 
closer and closer to the exact result. For 
$m_{H^\pm} \sim M_{\rm SUSY}$, the improvement provided by additional 
terms stops being so evident, at least in the case of the coefficient
$C_8(\mu_W)$.  Going further, it becomes quite clear that the results
obtained from the HME, and from the effective-Lagrangian formalism,
cannot be extended to values $m_{H^\pm} \gtap M_{\rm SUSY}$ (i.e.
strictly speaking outside the range in which they had been derived), 
as one may have hoped. For these values of $m_{H^\pm}$, the 
series of corrections in inverse powers of $M_{\rm SUSY}^2$ does not 
seem to be convergent.

In the safe region $m_{H^\pm} < M_{\rm SUSY}$, (that is, safe for the 
HME), we find that the largest contribution to $C_7(\mu_W)$ comes 
from the diagram on the left side of Fig.~\ref{CH-NLO}, with the 
photon emitted by the $t$-quark, for $C_8(\mu_W)$ from the two 
diagrams still on the left side of Fig.~\ref{CH-NLO}, with the gluon 
emitted by the $t$-quark and the gluino. This is true for the exact 
calculation and for the approximations at different order in 
$1/M_{\rm SUSY}^2$. Above this range of values for $m_{H^\pm}$, at
different orders of approximation, the contributions from different
diagrams tend to grow differently with $m_{H^\pm}$, producing the
crossing points of different lines, visible in Fig.~\ref{c78figlight}
and partially in Fig.~\ref{c78figheavy}.  For example, in the
approximation in which all terms 
${\cal O}(m_t^2,m_{H^\pm}^2/M_{\rm SUSY}^2)$ are retained, for a
specific value of $m_{H^\pm}$, all terms in $1/M_{\rm SUSY}^2$ cancel
out and the value of the Wilson coefficient coincides with that in which
only the nondecoupling terms are kept.  Similar cancellations of terms
happen in the other two crossing points.  In any case, there is no
particular meaning in these points, since they appear in a region in
which the HME is not well behaved.

For $m_{H^\pm}$ sufficiently smaller than $M_{\rm SUSY}$, the 
approximation in which only the nondecoupling terms of the 
two-loop diagrams are included, is not a bad approximation of the
exact calculation, for the coefficient $C_7(\mu_W)$. We observe, 
however, a 15\% deviation of the result of this approximation from 
that of the exact calculation for $C_8(\mu_W)$ in the case of the 
lighter supersymmetric spectrum. In this case, these two results 
seem to have a similar behaviour in $1/m_{H^\pm}^2$ for any value 
of $m_{H^\pm}$, but they are split by terms 
$(m_t^2/M_{\rm SUSY}^2)^n$, resummed in the exact calculation, and, 
possibly, by intrinsic constants arising from the two-loop 
calculation.

What is the impact of this deviation, and of our exact calculation in 
general, for the ${\rm BR}(\overline{B} \to X_s \gamma)$ and for an 
exclusion plot of the charged-Higgs mass remains to be seen. We have
not attempted to draw such a plot, since a calculation of the $W^\pm$
contribution at the same precision on $m_t^2/M_{\rm SUSY}^2$ is still
not yet available. Moreover, in the region of supersymmetric parameter
space in which the difference between our exact calculation and the
approximated one of Ref.~\cite{effVbsg1} is largest, one expect also a
rather big contribution from the chargino-stop exchange. We postpone the
presentation of such a plot to later work.

\section{Conclusion}
\label{concl}
We have studied the SUSY-QCD corrections to the charged-Higgs
contributions to the decay $\overline{B} \to X_s \gamma$. They are 
induced by gluino-squark subloops of 
${\cal O}(\alpha_s \tan \! \beta)$ and are therefore potentially 
large in the large-$\tan \! \beta$ regime.

The resulting two-loop diagrams had been dealt in the past in an 
approximate way.  In particular, they had been treated in the
approximation of an effective Lagrangian with two-Higgs doublets, in
which only the nondecoupling operators had been included.  For the
charged-Higgs contributions to $b\to s \gamma$ and $b \to s g$, this 
means that terms of ${\cal O}(m_t^2,m_{H^\pm}^2/M_{\rm SUSY}^2)$ or
higher, (with $M_{\rm SUSY}$ one of the typical squark/gluino masses)
are neglected.  Terms of this type are in general induced by
higher-dimensional operators. A truncation of the basis of operators in
the 2HD effective Lagrangian, is acceptable in particular directions of
parameter space, in which $m_{H^\pm}\ll M_{\rm SUSY}$. This condition,
however, is not generically supported by the most known and studied
models of supersymmetry breaking. In these, the charged-Higgs mass tends
to align along the gluino mass, in turn of the same order of the $\mu$
parameter.  (There are, however, directions in which a special tuning
among the different supersymmetric parameters allows considerably lower
values of $m_{H^\pm}$.)

For the charged-Higgs contributions to $b \to s \gamma$ and $b \to s g$, 
we have, therefore included all subleading terms, up to 
${\cal O}((m_t^2,m_{H^\pm}^2/M_{\rm SUSY}^2)^2)$, obtained through 
the technique of Heavy Mass Expansion of multi-loop integrals. This 
implicitly takes into account the contribution of all relevant 
higher-dimensional operators that should be added to the 
2HD effective Lagrangian. 
We have also performed the exact calculation of all two-loop diagrams
correcting at order ${\cal O}(\alpha_s \tan\!\beta)$ the charged-Higgs
contribution to $b \to s \gamma$ and $b \to s g$.  Thus, we have
compared to this exact result the different approximations obtained in
the HME, i.e. {\it 1)} that in which only the nondecoupling part of the
two-loop integrals is retained; {\it 2)} that in which terms of 
${\cal O}(m_t^2,m_{H^\pm}^2/M_{\rm SUSY}^2)$ are also added; and 
finally {\it 3)} that in which terms up to   
${\cal O}((m_t^2,m_{H^\pm}^2/M_{\rm SUSY}^2)^2)$ are included.

We have found that for $H^\pm$ considerably lighter than the remaining
supersymmetric particles, the result from the exact calculation and from
the three approximations deviate very little one from the other, for the
coefficient $C_7(\mu_W)$.  For the  coefficient $C_8(\mu_W)$ we find a 
deviation of 15\% between the result of 
approximation {\it 1)} and that of the exact calculation, if the 
supersymmetric spectrum is not
particularly heavy. 
The calculation presented here is a first step towards the exact
evaluation of all supersymmetric contributions to 
$\overline{B} \to X_s \gamma$ at order 
${\cal O}(\alpha_s \tan\! \beta)$. Only after the
completion of such a program, will phenomenological analyses be 
performed and exclusion plots for the masses of the
charged-Higgs and other supersymmetric particles be attempted.

\vspace{2mm}
{\it Acknowledgments:} 
The authors thank A.~Arhrib and J.L.~Kneur for checks with the
programs FeynArts~\cite{FEYNART} and SuSpect~\cite{SUSPECT},
respectively.  Thanks are also due to M.~Steinhauser for pointing
Ref.~\cite{GvdB} to us.
C.G. was supported by the Schweizerischer Nationalfonds. Y.Y. was
supported in part by the Grant-in-aid for Scientific Research from the
Ministry of Education, Culture, Sports, Science, and Technology of
Japan, No.~14740144.

\end{document}